\documentclass[10pt,letterpaper]{article}
\usepackage[OE]{express}
\usepackage{graphicx}
\usepackage{bm}
\usepackage{float}
\usepackage{ulem}
\usepackage{multicol}
\usepackage[utf8]{inputenc}
\newcommand{\sinc}{\ensuremath{\mbox{\hspace{0pt}sinc}}}

\begin{document}
% TITLE
\title{Tunable entanglement distillation of spatially correlated down-converted photons}
\author{E.~S.~G\'{o}mez,\authormark{1,*} P.~Riquelme,\authormark{2} M.~A.~Sol\'is-Prosser,\authormark{1,3} P.~Gonz\'alez,\authormark{1} E.~Ortega,\authormark{1,4} G.~B.~Xavier,\authormark{2,5} and G.~Lima\authormark{1,3}}
\address{\authormark{1}Departamento de F\'{i}sica, Universidad de Concepci\'{o}n, 160-C Concepci\'{o}n, Chile\\
\authormark{2}Departamento de Ingenier\'ia El\'ectrica, Universidad de Concepci\'on, 160-C Concepci\'on, Chile\\
\authormark{3}Millennium Institute for Research in Optics, Universidad de Concepci\'on, 160-C Concepci\'on, Chile.\\
\authormark{4}Currently with the Institute for Quantum Optics and Quantum Information (IQOQI), Austrian Academy of Sciences, Boltzmanngasse 3, 1090 Vienna, Austria\\
\authormark{5}Institutionen f\"{o}r Systemteknik, Link\"{o}pings Universitet, 581 83 Link\"{o}ping, Sweden}
\email{\authormark{*}estesepulveda@udec.cl}

% ABSTRACT
\begin{abstract}
We report on a new technique for entanglement distillation of the bipartite continuous variable state of spatially correlated photons generated in the spontaneous parametric down-conversion process (SPDC), where tunable non-Gaussian operations are implemented and the post-processed entanglement is certified in real-time using a single-photon sensitive Electron Multiplying CCD (EMCCD) camera. The local operations are performed using non-Gaussian filters modulated into a programmable spatial light modulator and, by using the EMCCD camera for actively recording the probability distributions of the twin-photons, one has fine control of the Schmidt number of the distilled state. We show that even simple non-Gaussian filters can be finely tuned to a $\sim$67\% net gain of the initial entanglement generated in the SPDC process.
\end{abstract}


\section{Introduction}
Entanglement represents correlations among quantum systems that cannot be explained by classical local models~\cite{ent}, and it is at the core of quantum information science. For instance, it is necessary for entanglement-based quantum cryptography \cite{ekert,Jene2000}, for teleportation of quantum states \cite{Bennet}, for quantum dense coding \cite{Bennet2}, and for Bell tests of quantum non-locality \cite{Bell,CHSH}. In photonic experiments, a suitable source of entangled photons is the spontaneous parametric down conversion (SPDC) process. SPDC is widely used to generate pairs of correlated photons that can be entangled in several degrees of freedom \cite{Mandel,PR}. For instance, the spatial entanglement of photons generated in the SPDC process has been used for fundamental studies of quantum mechanics~\cite{FonsecaBROGLIE,Tasca2012}, enhanced quantum imaging~\cite{Gatti,Brida,Boyd2}, and generation of high-dimensional quantum states~\cite{LeoGen,BoydQu,Zeil,obrien}.

Since the transverse momentum and position of single photons are continuous variables (CVs), one can use the spatially correlated down-converted photons for CV quantum information processing \cite{Braunstein,EPR,BoydEPR,Lantz}. Studies regarding the quantification of spatial entanglement in SPDC have been reported and one way to quantify the spatial entanglement is by using the Schmidt decomposition technique~\cite{Eberly,Fedorov1,Fedorov2,Just2013}. However, analytical computation of the Schmidt number $K$ is experimentally demanding since it requires a full state reconstruction \cite{SLM1,tomodardo}. The estimation of $K$ can be greatly simplified by noting that the spatial state of down-converted photons, under typical illumination procedure, is well approximated by a pure state \cite{Monken2}. In this case, $K$ can be obtained only by measuring the marginal distributions of the down-converted photons at the near-field (image) and far-field planes of the source. Therefore, the degree of spatial entanglement depends on the aforementioned distributions and, thus, modifications of them through the use of local spatial filters allow the entanglement manipulation \cite{filter}.

In this paper, we report on a novel experimental technique for entanglement distillation of the continuous variable (CV) spatial state of photons created in the SPDC process. Using tunable non-Gaussian local operations, and measuring in real time the marginal distributions, we developed a toolbox which allows a real-time, fine-tuning of the distilled entanglement. The filtering is achieved by using a spatial light modulator (SLM), which is a programmable liquid crystal display that can be used to control the amplitude and/or the phase of an incident beam. They have been used in several quantum information experiments~\cite{tomodardo,SLM1,HDQKD,SLM2}, showing the practicality for generating and manipulating $d$-dimensional quantum states encoded into the transverse linear momentum of single photons. The real-time recording of images is done by an Electron Multiplying Charge Coupled Device (EMCCD). The EMCCD camera is based on a technology consisting of a CCD image sensor capable of detecting single-photon level intensities with high frame rates, low readout noise, and with high quantum efficiency~\cite{LeoCam}. For this reason EMCCD cameras have been used in several recent quantum information and quantum imaging experiments \cite{Tasca2012,Boyd2,emccd2}.

Although SPDC is a widely employed source of entangled photons, the advantage of the combined use of tunable spatial filters and real-time recording of twin-photon marginal distributions for entanglement distillation has not been explored up to date. Our technique can find applications in recent applied research, where the propagation of spatially entangled photons through turbulent channels is considered. As the entanglement may deteriorate due to imperfections of the channel \cite{turb0,turb1,turb2,turb3}, the receiver might need to distillate the received state for quantum information processing. We also envisage that our technique can be used to generate entangled Bell states of orbital angular momentum modes with very high quantum numbers \cite{Torres1,OAMqd,OAMqd2,OAMqd3,OAMZ}.

\section{Spatial entanglement of the down-converted photons}
SPDC is a nonlinear optical process in which a laser beam with frequency $\omega_3$ and wave vector $\mathbf{k}_3$ pump a nonlinear birefringent crystal, in such a way that it generates two photons with frequencies $\omega_1$ and $\omega_2$ and wave vectors $\mathbf{k}_1$ and $\mathbf{k}_2$. If we assume the paraxial regime, a classical monochromatic continuous-wave pump beam, and the degenerate case, the spatial quantum state of the down-converted photons can be written as~\cite{PR,SPDC}

\begin{equation}\label{spdc_state}
|\Psi\rangle_{12}=\iint d\mathbf{q}_1 d\mathbf{q}_2\,v(\mathbf{q}_1+\mathbf{q}_2)g(\mathbf{q}_1-\mathbf{q}_2)|1\mathbf{q}_1\rangle|1\mathbf{q}_2\rangle.
\end{equation}
$\mathbf{q}_j$ represents the transverse momentum $\mathbf{q}$ for the j-th photon, namely $\mathbf{q}=\mathbf{k}-\mathbf{k}_z$. The function $v(\mathbf{q})$ is the angular spectrum of the pump beam. Typically, this function is a Gaussian distribution, namely $v(\mathbf{q})\propto\exp(-\sigma^2|\mathbf{q}|^2/4)$, where $\sigma$ is the beam waist. $g(\mathbf{q})\propto\sinc\frac{L}{4k_3}|\mathbf{q}|^2$ is the phase-matching function associated to the process, where $L$ is the crystal length, and $k_3=|\mathbf{k}_3|$ is the wavenumber of the pump beam. The phase-matching function implies that the state of Eq. (\ref{spdc_state}) is a non-Gaussian CV-state~\cite{tbo1,tbo2}. It can be rewritten in the transverse position domain through the Fourier transform as~\cite{Monken2,ExterNF2}
\begin{equation}\label{nf}
|\Psi\rangle_{12}=\iint d\mathbf{x}_1 d\mathbf{x}_2\,W\left(\frac{\mathbf{x}_1+\mathbf{x}_2}{2}\right)G\left(\frac{\mathbf{x}_1-\mathbf{x}_2}{2}\right)|1\mathbf{x}_1\rangle|1\mathbf{x}_2\rangle,
\end{equation}
where $W(\mathbf{x})$ and $G(\mathbf{x})$ are the Fourier transform of $v(\mathbf{q})$ and $g(\mathbf{q})$, respectively. The spatial CV-state given by Eqs. (\ref{spdc_state}) and (\ref{nf}) is entangled for any value of $\sigma$, $k_3$ and $L$ \cite{Eberly}.

In order to compute the degree of spatial entanglement, we use the Schmidt number $K$. This number is defined as the inverse of the purity of the marginal state $\rho_m$ of a bipartite state $\rho_{12}$, namely $K=1/\mathrm{Tr}(\rho_m ^2)$. However, a complete knowledge of the bipartite state $\rho_{12}$ is required to calculate the Schmidt number. Although this knowledge can be gathered through quantum state tomography, it would require one to perform many measurements in different bases to reconstruct the state~\cite{cvtomo}. Instead, we resorted to a recently proposed strategy to compute the Schmidt number associated to the spatial state of the down-converted photons~\cite{Monken2}. Assuming the state is pure, this strategy is based on measuring the probability distributions of the transverse momentum $\mathbf{q}_j$, and the transverse position $\mathbf{x}_j$, of one of the twin-photons associated to the state described by Eqs.~(\ref{spdc_state}) and ~(\ref{nf}). The Schmidt number is given by
\begin{equation}\label{Kmonken}
K=\frac{1}{4\pi^2}\frac{[\int d\mathbf{x}_j\,I(\mathbf{x}_j)]^2}{\int d\mathbf{x}_j\,I^2(\mathbf{x}_j)}\times\frac{[\int d\mathbf{q}_j\,I(\mathbf{q}_j)]^2}{\int d\mathbf{q}_j\,I^2(\mathbf{q}_j)},
\end{equation}
where the probability distributions for the $j$-th photon are
\begin{align}
I(\mathbf{q}_j) &= \int d\mathbf{q}_i\, v^2(\mathbf{q}_i+\mathbf{q}_j)g^2(\mathbf{q}_i-\mathbf{q}_j), \quad\mathrm{and}\label{ff1}\\
I(\mathbf{x}_j) &= \int d\mathbf{x}_i\, W^2\left(\frac{\mathbf{x}_i+\mathbf{x}_j}{2}\right)G^2\left(\frac{\mathbf{x}_i-\mathbf{x}_j}{2}\right).\label{nf1}
\end{align}
Under typical experimental conditions, i.e., considering a thin crystal and a wide pump beam, one has that $\sigma\gg \frac{L}{K}$. Thus, Eqs. (\ref{ff1}) and (\ref{nf1}) can be simplified by considering $v^2(\mathbf{q}_1+\mathbf{q}_2)\approx\delta(\mathbf{q}_1+\mathbf{q}_2)$, and $G^2\left(\frac{\mathbf{x}_1-\mathbf{x}_2}{2}\right)\approx\delta\left(\mathbf{x}_1-\mathbf{x}_2\right)$, and one obtains that
\begin{align}
I(\mathbf{q}) &= g^2(2\mathbf{q})\propto\sinc^2\left(\frac{L}{K}|\mathbf{q}|^2\right), \quad\mathrm{and}\label{ff2}\\
I(\mathbf{x}) &= W^2(\mathbf{x})\propto\exp\left(-\frac{2|\mathbf{x}|^2}{\sigma^2}\right).\label{nf2}
\end{align}
Note that due to the symmetry of the state, the probability distributions are the same for both down-converted photons. Therefore, the Schmidt number $K$ is given by
\begin{equation}\label{K0}
K=\frac{3\pi^2\sigma^2}{8\lambda_3 L}.
\end{equation}

Since the spatial entanglement depends on the widths and amplitude of both marginal probability distributions, it is possible to manipulate the amount of entanglement by adding a spatial filter at the image plane of the source for both down-converted photons~\cite{filter}. Considering a local transmission function $F(\mathbf{x})$, the two-photon state after the spatial filter is given by
\begin{equation}\label{Eqfiltered}
|\Psi\rangle_{12}^{(F)}=\iint d\mathbf{x}_1 d\mathbf{x}_2\,W\left(\frac{\mathbf{x}_1+\mathbf{x}_2}{2}\right)F(\mathbf{x}_1)F(\mathbf{x}_2)G\left(\frac{\mathbf{x}_1-\mathbf{x}_2}{2}\right)|1\mathbf{x}_1\rangle|1\mathbf{x}_2\rangle.
\end{equation}
Depending on the shape of the filter, we can increase the spatial entanglement, despite losing part of the photons sent. This is the protocol known as entanglement distillation. In our case, the spatial filter is implemented using a programmable SLM, and the modified marginal probability distributions are recorded in real time using the EMCCD camera.

Please note there are other approaches to certifying the spatial entanglement of down-converted photons such as the use of separability criteria for entangled CV-states \cite{reid,luan,mancini}. Nonetheless, in the case of spatially correlated SPDC photons, these criteria require the coincidence rate measurements in two conjugate planes and this is experimentally demanding \cite{Eisert,steveng}.

\section{Experiment}
\begin{figure*}[t]
\centering
\includegraphics[width=0.98\textwidth]{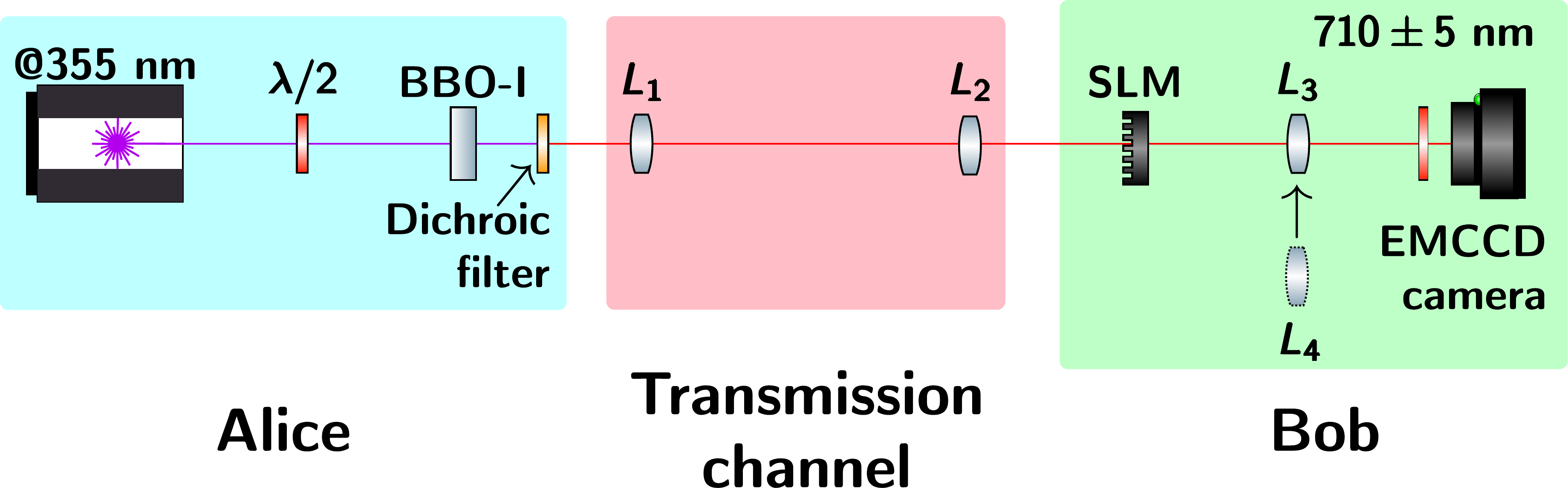}
\caption{Experimental setup. See the main text for more details.\label{fig-exp} }
\end{figure*}

The experimental setup is depicted in Fig.~\ref{fig-exp}. Alice uses a continuous-wave, single-mode laser operating at 355 nm to pump, a $\beta$-barium-borate type-I (BBO-I) nonlinear crystal to generate collinear down-converted photons. The maximum laser power is 350 $mW$. The nominal pump beam waist is 600 $\mu m$, and the crystal length is 4.5 mm. After the BBO crystal, high-quality dichroic mirrors are used to remove the remaining pump beam, transmitting only the down-converted photons. The generated photons are sent to Bob through the transmission channel. In our case, this channel corresponds to a free space separation between Alice and Bob, composed of two lenses, $L_1$ and $L_2$, both with 7.5 cm of focal length, in a 4-$f$ configuration.

To test our entanglement distillation protocol, Bob uses a transmissive spatial light modulator (SLM) working in amplitude-mode to perform the spatial filtering. The SLM is composed of an array of $800\times 600$ pixels (px), which are 32 $[\mu m]$ wide, sandwiched between cross-aligned polarizers. Bob places the SLM at the propagation mode of both down-converted photons. Even though it acts on both photons, the applied spatial filter is local because the total transmission function is given by $F(\mathbf{x}_1,\mathbf{x}_2)=F(\mathbf{x}_1)F(\mathbf{x}_2)$. This filter allows the manipulation of the probability distribution of the transverse position of $\mathbf{x}_j$ since it is located at the image plane of the source (see Fig.~\ref{fig-exp}). As it is well known, the modulation of the intensity on the near-field plane of the BBO-I crystal corresponds to a modification on the transverse position distributions of the twin-photons \cite{ExterNF2,BoydEPR}.

Even though the spatial state of the down-converted photons is non-Gaussian~\cite{tbo1,tbo2}, it can be approximated by a Gaussian one~\cite{PR,Boyd2,BoydEPR,Eberly}. Therefore, it is natural to consider a non-Gaussian spatial filter since it is impossible to implement an entanglement distillation protocol under Gaussian states using Gaussian operations~\cite{Gaussimp}.  Thus, we consider the local non-Gaussian filter given by
\begin{equation}\label{ngfilter}
F(\mathbf{x})=F(x,y)=\left[e^{-\frac{(x-d)^2}{4a^2}}+e^{-\frac{(x+d)^2}{4a^2}}\right]\cdot\left[e^{-\frac{(y-d)^2}{4a^2}}+e^{-\frac{(y+d)^2}{4a^2}}\right].
\end{equation}
This function, for each transverse direction $x,y$, is composed of a sum of two Gaussian functions, where $\pm d$ is their offset position, and $a$ is their width. The filter given by Eq.~(\ref{ngfilter}) represents a family of non-Gaussian filters depending on the values of~$d$ and~$a$. Then, by finely tuning the values of $d$ and/or $a$ of the filter addressed in the SLM, one obtains different post-selected entangled states.

To experimentally monitor in real-time the manipulation of the spatial entanglement after the photons cross the filter, an EMCCD camera is used to record the marginal probability distribution of the transverse position and momentum of a down-converted photon. This measurement is done at the image plane (near-field) and the Fourier plane (far-field) of the filter, respectively. In particular, we used a Princeton ProEM 512 camera, with a resolution of $512\times 512$ px. Each camera's pixel has a width equal to 16 $\mu m$. In the case of measuring at the near-field plane, we resort to the use of the lens $L_3$ with a 7.5 cm of focal length. When $L_4$ (15 cm focal length) is placed instead of $L_3$, the camera is located at the Fourier plane of the SLM. An interference filter placed at the front of the EMCCD camera was used to select the degenerated down-converted photons at a wavelength of 710 nm with 5 nm of bandwidth.

It is worth to note that in our scheme the purity of the post-selected states is guaranteed such that Eq. (\ref{Kmonken}) is still valid. For instance, decoherence effects generating a mixed spatial state would arise if the single-photon spatial degree of freedom entangles with other degrees of freedom, e.g. polarization, time-bin, etc. Nevertheless, in our case this is avoided by using polarization filters, spectral filters, and the repetition rate of the SLM in large timescales compared with the coherence time of the down-converted photons~\cite{Lizana2008}. Thus, avoiding spatial-polarization, spatial-frequency and spatial-temporal correlations, respectively.

\section{Results}
To compute the Schmidt number of the entangled generated states, we initially fit a surface to each image recorded by the EMCCD at the near- and far-field planes. According to the state given by Eq.~(\ref{Eqfiltered}), and the filter function of Eq.~(\ref{ngfilter}), the fitting surface must be a sum of four Gaussian functions for the near-field plane images
\begin{equation}\label{fitnf}
I^{NF}_{fit}(x,y)=\sum_{i=1}^4 \alpha_i\exp\left(-\frac{(x-\beta_i)^2}{\delta_i^2}\right)\exp\left(-\frac{(y-\gamma_i)^2}{\epsilon_i^2}\right),
\end{equation}
and a sinc function for the far-field plane images
\begin{equation}\label{fitff}
I^{FF}_{fit}(x,y)=a+b\,\sinc^2\left(c\left[(x-x_0)^2+(y-y_0)^2\right]\right).
\end{equation}
Each integral in Eq. (\ref{Kmonken}) is calculated based on the obtained parameters of the fitted surfaces (i.e. $\alpha_i,\beta_i,\gamma_i,\delta_i,\epsilon_i,a,b,c,x_0,$ and $y_0$). That is, the Schmidt number $K$ for the post-selected states can be calculated directly using these parameters.

\subsection{Measuring the initial spatial entanglement}
\begin{figure}[!t]
\centering
\includegraphics[width=0.9\textwidth]{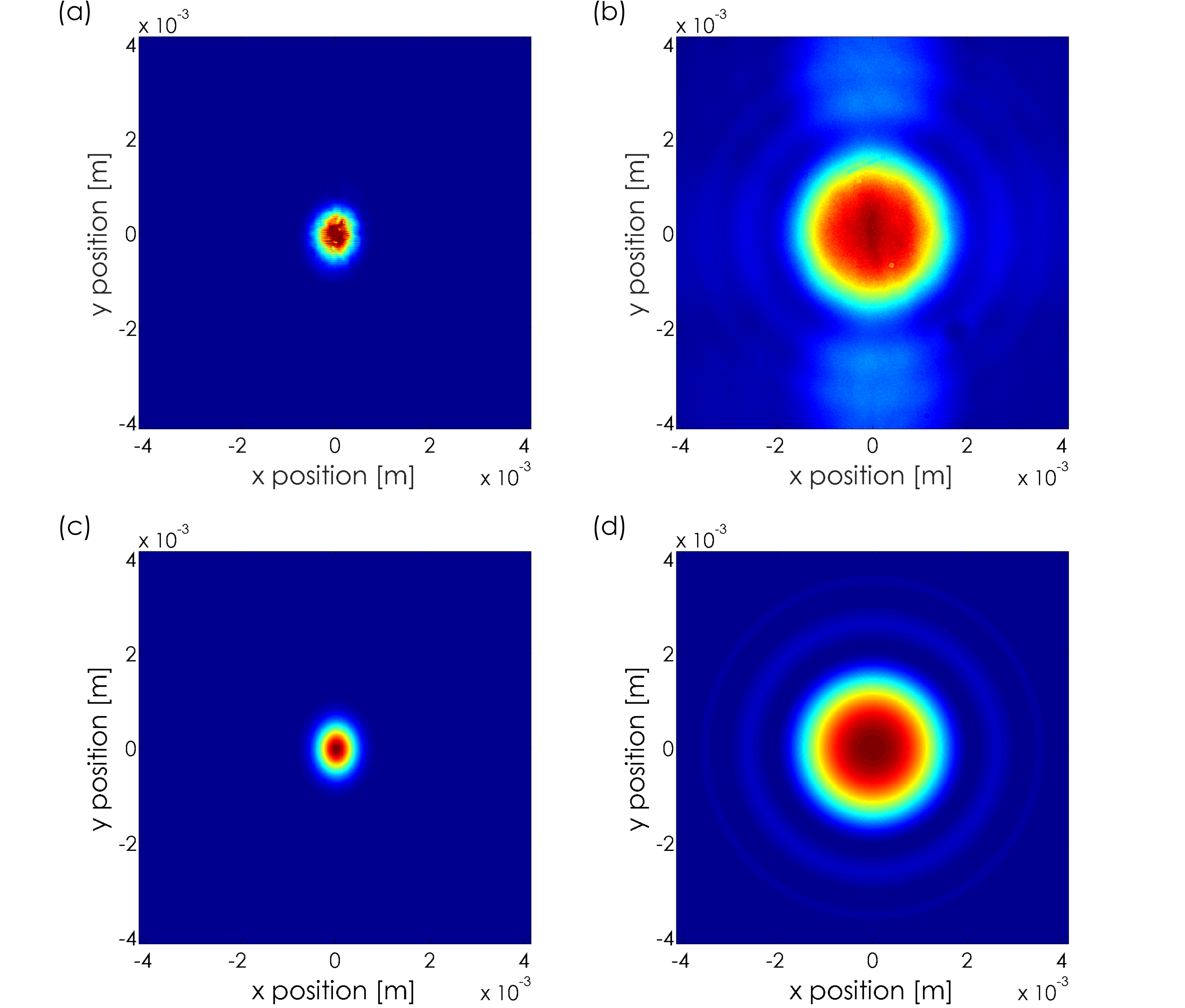}
\caption{Images recorded with the EMCCD camera for computing the initial Schmidt number $K_0$. In (a) we show the near-field distribution generated by the transverse pump beam. In (b) we show the far-field distribution that arise from the phase-matching conditions. (c) and (d) show the corresponding fitting surfaces.\label{fig:K0}}
\end{figure}

To properly estimate the efficiency of our technique for the spatial entanglement distillation, we started by measuring the initial spatial entanglement of the down-converted photons generated at the crystal. For this purpose no filter was addressed in the SLM, and it was set to a ``blank configuration'' corresponding to the maximal transmittance setting. The recorded images and the fitted ones are shown in Fig.~\ref{fig:K0}. The presence of an unwanted vertical diffraction pattern in the recorded image at the far-field plane can be spotted. It is caused by non-modulated light scattered on periodic pixel structure of the SLM and, therefore, it corresponds to background noise and does not affect/impact on the estimation of the degree of entanglement of the distilled states.

From the recorded images, we obtain that the initial Schmidt number is $K_0=842\pm40$. The error was extracted from the computational routine which gives the goodness of the fitted surfaces. This result is in a good agreement with the predicted Schmidt number of $K_{theo}=834.05$, obtained considering the experimental values of $\sigma$, $L$, and $\lambda_3$ into Eq.~(\ref{K0}).

\begin{figure}[!b]
\centering
\includegraphics[width=1\textwidth]{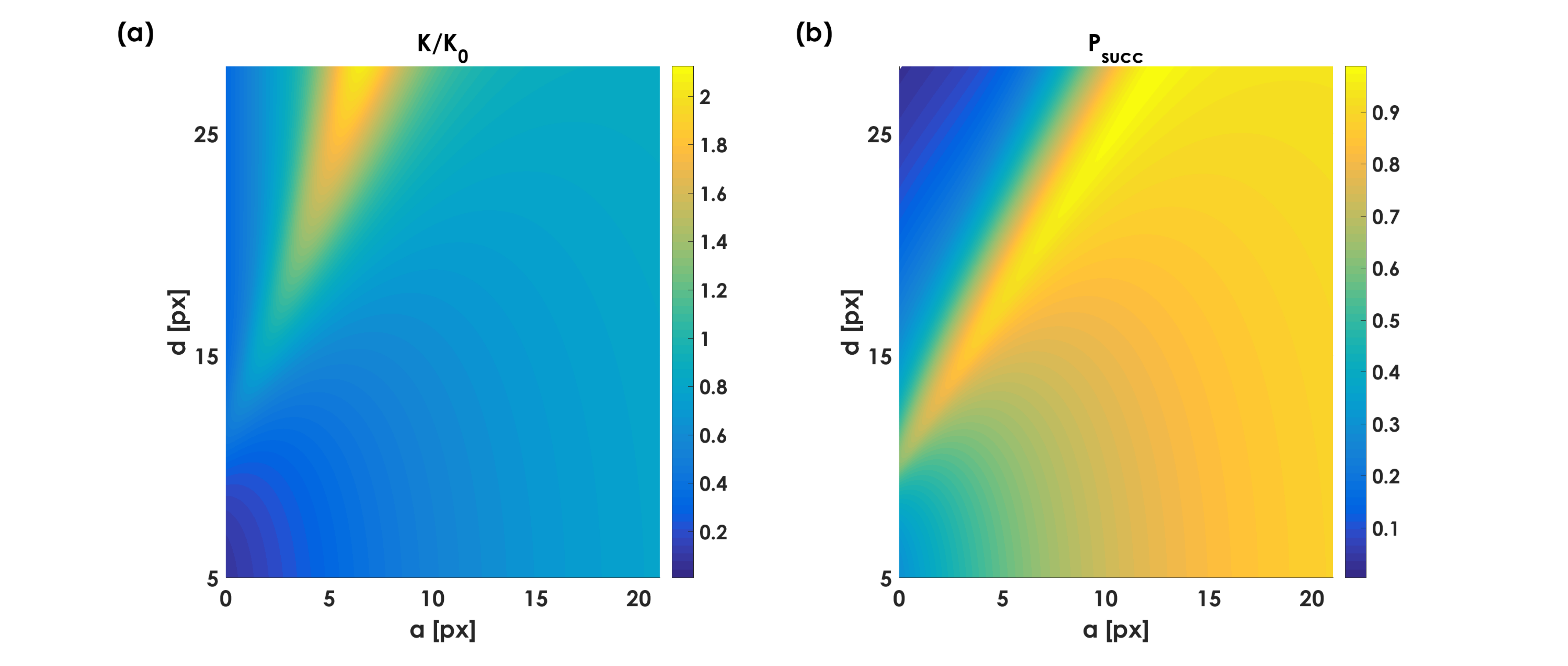}
\caption{Numerical simulations for our entanglement distillation protocol. (a) The ratio $K/K_0$ of the post-selected states when $a$ and $d$ are varying.  (b) The success probability $P_{\text{succ}}$ as function of $a$ and $d$. \label{fig:sim}}
\end{figure}

\subsection{Experimental spatial entanglement distillation}
\begin{figure}[!t]
\centering
\includegraphics[width=0.9\textwidth]{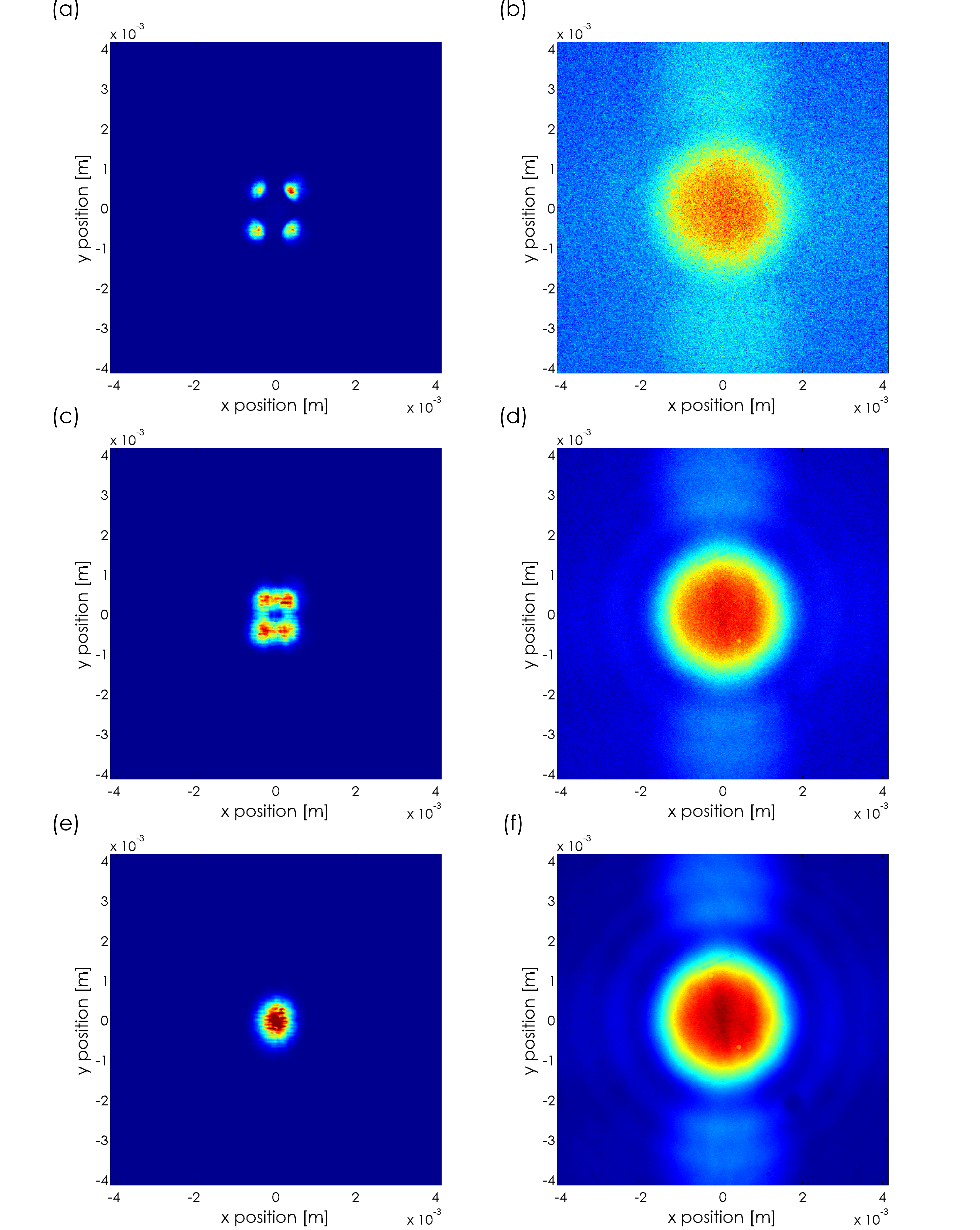}
\caption{Some examples of the experimental results obtained for the first scenario considered, where $d$ is fixed to 20 px and $a$ is varying. In (a), (c) and (e) we show the recorded images at the near-field plane when $a=6,8$ and 14 px, respectively. (b), (d) and (f) show the corresponding recorded images at the far-field plane when $a=6,8$ and 14 px, respectively.\label{fig:res1}}
\end{figure}

\begin{figure}[!t]
\centering
\includegraphics[width=0.9\textwidth]{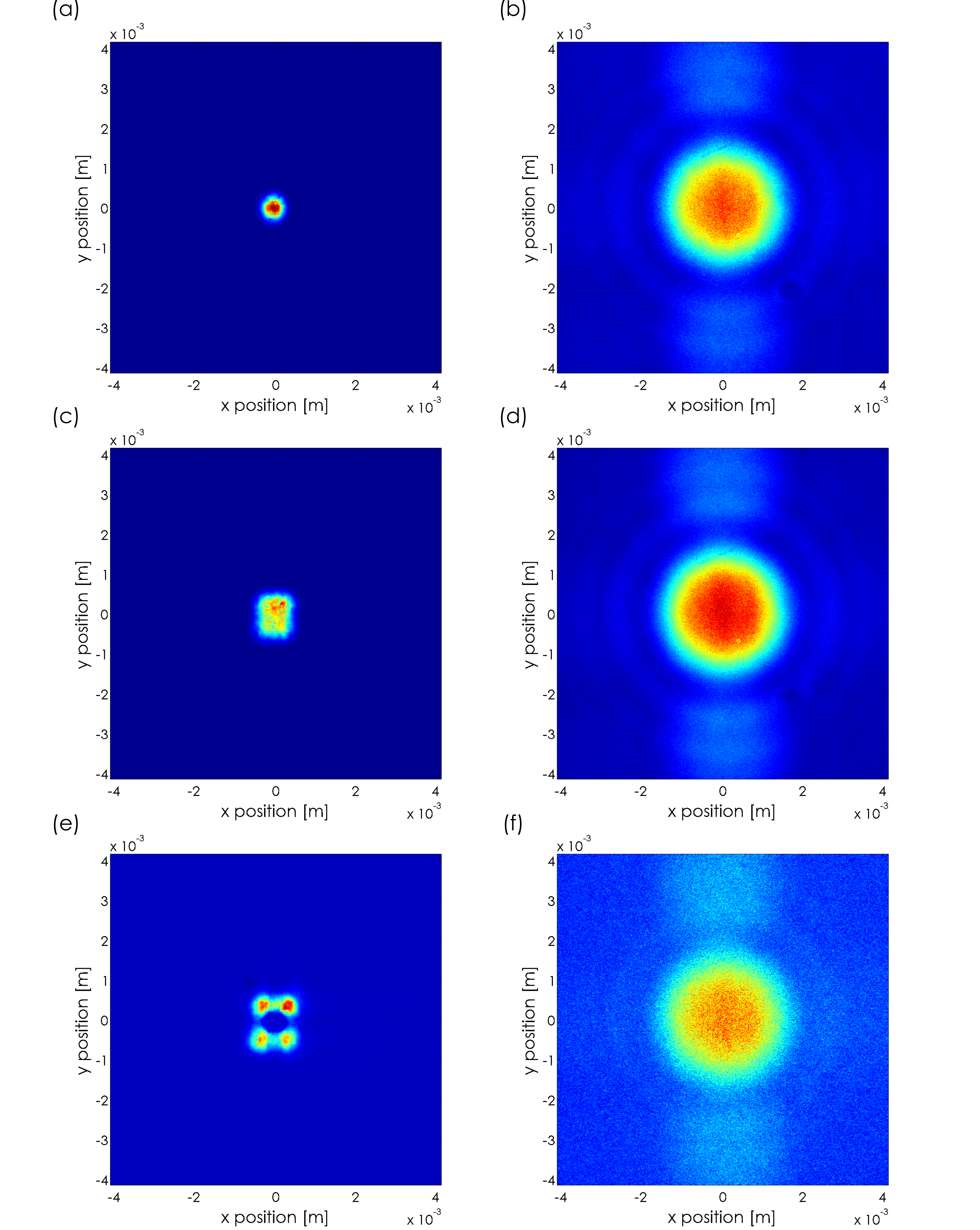}
\caption{Some experimental results obtained for the second scenario considered, where $d$ is varying. In (a), (c) and (e) we show the recorded images at the near-field plane when $d=$0, 14 and 19 px, respectively. In (b), (d) and (f) we show the corresponding recorded images at the far-field plane when $d=0,14$ and 19 px, respectively.\label{fig:res2}}
\end{figure}

We consider two main scenarios with different spatial filter configurations for the spatial entanglement distillation protocol: (i) in the first case, $d$ is fixed at 20 px while $a$ varies from 4 to 21 px, and (ii) $d$ varies between 0 and 21 px while $a$  is fixed at 7 px. These arbitrary values for $d$ and $a$ were chosen after simulating the effect of the non-Gaussian filters and noting that the post-selected entangled states would in some cases show high values of $K$ such that it is greater than $K_0$. The simulation results are shown in Fig. \ref{fig:sim}. In Fig. \ref{fig:sim}(a) we show the ratio $K/K_0$ of the post-selected states when $a$ and $d$ are varying. In Fig. \ref{fig:sim}(b) we show the success probability $P_{\text{succ}}$ of the protocol. It is defined as the probability for the transmission of the down-converted photons through the filter addressed in the SLM, conditioned on their initial state of Eqs. (\ref{spdc_state}) and (\ref{nf}). Some experimental results are shown in Figs.~\ref{fig:res1} and \ref{fig:res2}. Specifically, Fig.~\ref{fig:res1} shows the recorded distributions in the near- and far-field planes of the SLM for three cases belonging to the first scenario, where $a$ is varying. The first case corresponds to $a=6$ px, and the images of the near- and far-field planes are shown in Figs. \ref{fig:res1}(a) and \ref{fig:res1}(b), respectively. Figs. \ref{fig:res1}(c) and \ref{fig:res1}(d) show the case of $a=8$ px, corresponding to the maximum amount of distilled entanglement in our experiment for the first scenario. The third case of Figs. \ref{fig:res1}(e) and \ref{fig:res1}(f) corresponds to $a=14$ px, thus showing the overall behavior of such probability distributions for the first scenario.

Fig.~\ref{fig:res2} shows the probability distributions associated to 3 other cases belonging to the second scenario considered, where $d$ is varying. The first case corresponds to $d=0$ px, and the recorded images of the near- and far-field planes are shown in Figs. \ref{fig:res2}(a) and \ref{fig:res2}(b), respectively. Figs. \ref{fig:res2}(c) and \ref{fig:res2}(d) show the case of $d=14$ px. The third case depicted in Figs. \ref{fig:res2}(e) and \ref{fig:res2}(f) corresponds to $d=19$ px, thus also showing the overall behavior of such probability distributions.

\begin{figure}[t]
\centering
\includegraphics[width=1\textwidth]{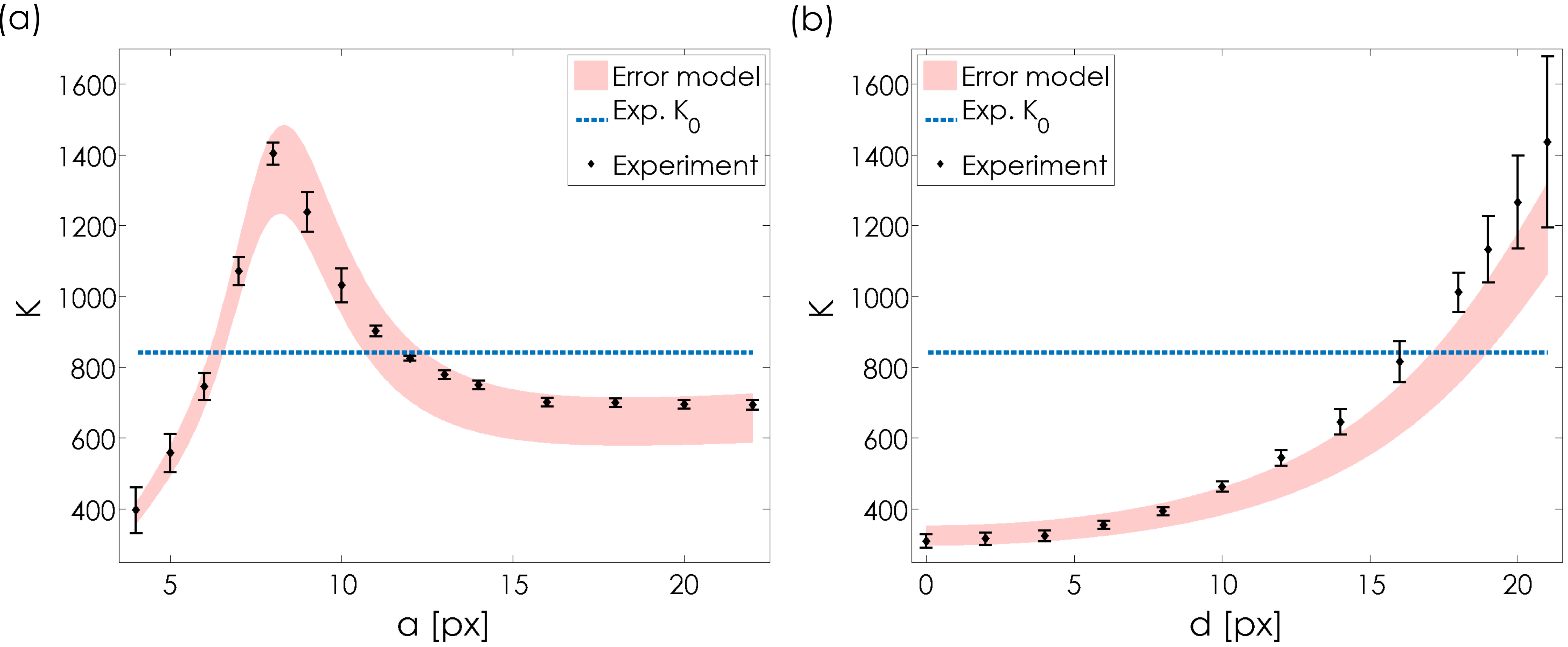}
\caption{Experimental results for both scenarios considered for the entanglement distillation procedure. In (a) we show the recorded $K$ values when $d$ is fixed at 20 $px$ while $a$ is varying. In (b) we show the obtained $K$ values when $a$ is fixed at 7$px$ and $d$ is varying. Black dots correspond to the experimental results and the blue dashed line shows the initial value of the spatial entanglement $K_0$. The pink area represents the confidence bound of $K$ obtained by Monte-Carlo simulations and based on the experimental errors.  \label{fig:var_a}}
\end{figure}

\begin{table}[!h]
\begin{center}
\caption{Experimental results for $K$ when $d$ is fixed at 20 $px$ while $a$ is varying.
\label{table-vara}}
\begin{tabular}{ccc}
\hline
$a$ (px) & $K$ & $\Delta K$ \\ \hline
$4$  & $397$ & $65$\\
$5$ & $558$ & $54$\\
$6$ & $745$ & $38$\\
$7$ & $1071$ & $40$\\
$8$ & $1404$ & $31$\\
$9$ & $1238$ & $56$\\
$10$ & $1031$ & $48$\\
$11$ & $902$ & $15$\\
$12$ & $825$ & $7$\\
$13$ & $779$ & $12$\\
$14$ & $753$ & $12$\\
$16$ & $717$ & $12$\\
$18$ & $699$ & $12$\\
$20$ & $695$ & $12$\\
$22$ & $694$ & $13$\\
\hline
\end{tabular}
\end{center}
\end{table}

\begin{table}[!h]
\begin{center}
\caption{Experimental results for $K$ when $a$ is fixed at 7 $px$ while $d$ is varying.\label{table-vard}}
\begin{tabular}{ccc}
\hline
$d$ (px) & $K$ & $\Delta K$ \\ \hline
$0$  & $309$ & $ \ \ 19$ \\
$2$  & $316$ & $ \ \ 18$ \\
$4$  & $324$ & $ \ \ 15$ \\
$6$  & $355$ & $ \ \ 12$ \\
$8$  & $393$ & $ \ \ 12$ \\
$10$ & $463$ & $ \ \ 15$ \\
$12$ & $544$ & $ \ \ 22$ \\
$14$ & $646$ & $ \ \ 36$ \\
$16$ & $815$ & $ \ \ 58$ \\
$18$ & $1011$ & $ \ \ 56$ \\
$19$ & $1132$ & $ \ \ 94$ \\
$20$ & $1266$ & $131$ \\
$21$ & $1435$ & $242$ \\ \hline
\end{tabular}
\end{center}
\end{table}

As we mentioned before, our technique is based on tunable spatial filters addressed in programmable SLMs to finely tune the distilled entanglement of the post-selected states. For the first scenario considered, we gradually generate 15 new states by only varying $a$. For the second scenario where $d$ is varying, we continuously generate 13 new states. The Schmidt numbers were computed for each configuration adopted. The corresponding results for the first scenario are shown in Fig.~\ref{fig:var_a}(a), and explicitly given in Table \ref{table-vara}. The obtained results related to the second scenario are shown in Fig.~\ref{fig:var_a}(b) and given in Table \ref{table-vard}. The dashed blue line in Figs.~\ref{fig:var_a}(a) and \ref{fig:var_a}(b) corresponds to the initial entanglement $K_0$. The black dots correspond to the experimental results obtained with different non-Gaussian filters. The pink areas represent confidence bounds computed using the Monte Carlo method and the uncertainty of $\sigma$ and $L$ values. As it was mentioned above, $\sigma=600\,\mu m$ and $L=4.5$ mm. The corresponding errors are $\Delta\sigma=13\,\mu m$ and $\Delta L=0.3$ mm, and they are consequence of systematic errors in the used measuring devices. Therefore, in the Monte Carlo method the computation of $K$, for each non-Gaussian filter, was done by randomly adopting $\sigma$ and $L$ values within $\sigma\pm\Delta\sigma$ and $L\pm\Delta L$. Please note the agreement between the experimental data and the expected values within the confidence bounds.  In Fig.~\ref{fig:var_a}(a) one can observe an increment of the initial entanglement when $6<a<12$ px. The most entangled post-selected state is obtained when $a=8$~px. In this case, $K=1404\pm31$, which corresponds to a $\sim$67\% net gain of the initial measured spatial entanglement.

Fig.~\ref{fig:var_a}(b) shows that with $a=7$~px entanglement distillation occurs when $d\geq18$~px, with no apparent upper bound within the observed range. The reason why larger values of $d$ were not tested lies in the fact that the probability of the photons to successfully pass through the filter becomes smaller, making the number of detected photons smaller as well. In turn, a smaller number of detected photons produces poorer signal-to-noise ratios, as evidenced in the larger error bars of the experimental data as $d$ grows. This observation suggests that keeping a constant value for $d$ while searching for a good value of $a$ yields better results for the entanglement distillation than the other way around.

\section{Conclusion}
We report on a toolbox technique that allows one to increase the amount of spatial entanglement of the down-converted photons. It is based on the implementation of tunable local non-Gaussian filters encoded onto programmable spatial light modulators, while the post-processed distilled entanglement is measured in real-time using an EMCCD single-photon sensitive camera to register the probability distributions of the transverse position and momentum of the down-converted photons. We showed that by finely tuning simple non-Gaussian filters one can already obtain a $\sim$67\% net gain of the initial spatial entanglement generated at the SPDC process. As entanglement is the core resource for quantum information processing, this procedure should find applications in several different scenarios. For instance, in quantum cryptography schemes where OAM correlated photons are propagated through the turbulent atmosphere and then used to establish a secret key for secure communications \cite{QKDOAM1,QKDOAM2}. Other CV schemes, such as computation based on continuous-variable cluster states might also benefit from this technique \cite{Zhang06,Menicucci06,Yukawa08}.

%%%%%%%%%%%%%%%%%%%%%%%%%%%%%%%%%%%%%%%%%%%%%%%%%%%%%%%%%%%%%%%%%%%%%%%%%%%%%

%%%%%%%%%%%%%%%%%%%
\section*{Funding}
Fondecyt (11150325, 1160400, 3170400, 1150101); PAI-Conicyt (79160083); Millennium Institute for Research in Optics (MIRO); Becas CONICYT.


\begin{thebibliography}{999}
\bibitem{ent} R. Horodecki, P. Horodecki, M. Horodecki, and K. Horodecki, ``Quantum entanglement,'' \rmp \textbf{81}, 865 (2009).

\bibitem{ekert} A. Ekert, ``Quantum cryptography based on Bell’s theorem,'' \prl~\textbf{67}, 661 (1991) .

\bibitem{Jene2000}  T. Jennewein, C. Simon, G. Weihs, H. Weinfurter, and A. Zeilinger, ``Quantum cryptography with entangled photons,'' \prl~\textbf{84}, 4729 (2000).

\bibitem{Bennet} C. H. Bennett, G. Brassard, C. Crépeau, R. Jozsa, A. Peres, and W. K. Wootters, ``Teleporting an unknown quantum state via dual classical and Einstein-Podolsky-Rosen channels,'' \prl~\textbf{70}, 1895 (1993).

\bibitem{Bennet2} C. H. Bennett and S. J. Wiesner, ``Communication via one- and two-particle operators on Einstein-Podolsky-Rosen states,'' \prl \textbf{69}, 2881 (1992).

\bibitem{Bell}  J. Bell, ``On the Einstein Podolsky Rosen paradox,'' Physics \textbf{1}, 195 (1964).

\bibitem{CHSH} J. F. Clauser, M. A. Horne, A. Shimony, and R. A. Holt, ``Proposed experiment to test local hidden variable theories,'' \prl \textbf{24}, 549 (1970).

\bibitem{Mandel} C. K. Hong and L. Mandel, ``Theory of parametric frequency down conversion of light,'' \pra~\textbf{31}, 2409 (1985).

\bibitem{PR} S. P. Walborn, C. H. Monken, S. P\'{a}dua, and P. H. S. Ribeiro, ``Spatial correlations in parametric down-conversion,'' Physics Reports \textbf{495}, 87 (2010).

\bibitem{FonsecaBROGLIE} E. J. S. Fonseca, C. H. Monken, and S. P\'{a}dua, ``Measurement of the de Broglie wavelength of a multiphoton wave packet,'' \prl~\textbf{82}, 2868 (1999).

\bibitem{Tasca2012} M.P. Edgar, D.S. Tasca, F. Izdebski, R.E. Warburton, J. Leach, M. Agnew, G.S. Buller, R.W. Boyd, and M.J. Padgett, ``Imaging high-dimensional spatial entanglement with a camera,'' Nat. Comm. \textbf{3}, 984 (2012).

\bibitem{Gatti} A. Gatti, E. Brambilla, and L. A. Lugiato, ``Entangled imaging and wave-particle duality: from the microscopic to the macroscopic realm,'' \prl~\textbf{90}, 133603 (2003).

\bibitem{Brida} G. Brida, M. Genovese, and I. Ruo Berchera, ``Experimental realization of sub-shot-noise quantum imaging,'' Nat. Phot. \textbf{4}, 227-230 (2010).

\bibitem{Boyd2} P. A. Morris, R. S. Aspden, J. E. C. Bell, R. W. Boyd, and M. J. Padgett, ``Imaging with a small number of photons'', Nat. Comm. \textbf{6}, 5913 (2015).

\bibitem{LeoGen} L. Neves, G. Lima, J. G. Aguirre G\'omez, C. H. Monken, C. Saavedra, and S. P\'adua, ``Generation of entangled states of qudits using twin photons,'' \prl~\textbf{94}, 100501 (2005).

\bibitem{BoydQu} M. N. O'Sullivan-Hale, I. Ali Khan, R. W. Boyd, and J. C. Howell, ``Pixel entanglement: experimental realization of optically entangled $d=3$ and $d=6$ qudits,'' \prl~\textbf{94}, 220501 (2005).

\bibitem{Zeil} M. Krenn, M. Malik, M. Erhard, and A. Zeilinger, ``Orbital angular momentum of photons and the entanglement of Laguerre-Gaussian modes,'' Phil. Trans. R. Soc. A \textbf{375} 20150442 (2017).

\bibitem{obrien} J. Wang, S. Paesani, Y. Ding, R. Santagati, P. Skrzypczyk, A. Salavrakos, J. Tura, R. Augusiak, L. Man\v{c}inska, D. Bacco, D. Bonneau, J. W. Silverstone, Q. Gong, A. Acín, K. Rottwitt, L. K. Oxenl{\o}we, J. L. O'Brien, A. Laing, and M. G. Thompson, ``Multidimensional quantum entanglement with large-scale integrated optics,'' Science \textbf{360}, 285 (2018).

\bibitem{Braunstein}  S. Braunstein and P. van Lock, ``Quantum information with continuous variables,'' Rev. Mod. Phys. \textbf{77}, 513 (2005).

\bibitem{EPR} A. Einstein, B. Podolsky and N. Rosen, ``Can quantum-mechanical description of physical reality be considered complete?,'' Phys. Rev. \textbf{47}, 777 (1935).

\bibitem{BoydEPR} J. C. Howell, R. S. Bennink, S. J. Bentley, and R. W. Boyd, ``Realization of the Einstein-Podolsky-Rosen paradox using momentum- and position-entangled photons from spontaneous parametric down conversion,'' \prl~\textbf{92}, 210403 (2004).

\bibitem{Lantz} P. -A. Moreau, F. Devaux, and E. Lantz, ``Einstein-Podolsky-Rosen paradox in twin images,'' \prl~\textbf{113}, 160401 (2014).

\bibitem{Eberly} C. K. Law and J. H. Eberly, ``Analysis and interpretation of high transverse entanglement in optical parametric down conversion,'' \prl~\textbf{92}, 127903 (2004).

\bibitem{Fedorov1} M. V. Fedorov, M. A. Efremov, A. E. Kazakov, K. W. Chan, C. K. Law, and J. H. Eberly, ``Packet narrowing and quantum entanglement in photoionization and photodissociation,'' Phys. Rev. A \textbf{69}, 052117 (2004).

\bibitem{Fedorov2}
G. Brida, V. Caricato, M. V. Fedorov, M. Genovese, M. Gramegna, and S. P. Kulik, ``Characterization of spectral entanglement of spontaneous parametric-down conversion biphotons in femtosecond pulsed regime,'' EPL \textbf{87}, 64003 (2009).

\bibitem{Just2013} F. Just, A. Cavanna, M. V. Chekhova, and G. Leuchs, ``Transverse entanglement of biphotons, ``New J. Phys. \textbf{15}, 083015 (2013).

\bibitem{SLM1}  G. Lima, L. Neves, R. Guzmán, E. S. G{\'o}mez, W. A. T. Nogueira, A. Delgado, A. Vargas, and C. Saavedra, ``Experimental quantum tomography of photonic qudits via mutually unbiased basis,'' Opt. Express \textbf{19}, 3542 (2011).

\bibitem{tomodardo} D. Goyeneche, G. Cañas, S. Etcheverry, E. S. Gómez, G. B. Xavier, G. Lima, and A. Delgado, `` Five measurement bases determine pure quantum states on any dimension,'' \prl \textbf{115}, 090401 (2015).

\bibitem{Monken2} H. Di Lorenzo Pires, C. H. Monken, and M. P. van Exter, ``Direct measurement of transverse-mode entanglement in two-photon states,'' \pra~\textbf{80}, 022307 (2009).

\bibitem{filter} M. P. van Exter, A. Aiello, S. S. R. Oemrawsingh, G. Nienhuis, and J. P. Woerdman, ``Effect of spatial filtering on the Schmidt decomposition of entangled photons,'' \pra~\textbf{74}, 012309 (2006).

\bibitem{HDQKD} G. Cañas, N. Vera, J. Cariñe, P. González, J. Cardenas, P. W. R. Connolly, A. Przysiezna, E. S. Gómez, M. Figueroa, G. Vallone, P. Villoresi, T. Ferreira da Silva, G. B. Xavier, and G. Lima, ``High-dimensional decoy-state quantum key distribution over multicore telecommunication fibers,'' \pra \textbf{96}, 022317 (2017).

\bibitem{SLM2}  G. Cañas, S. Etcheverry, E. S. Gómez, C. Saavedra, G. B. Xavier, G. Lima, and A. Cabello, ``Experimental implementation of an eight-dimensional Kochen-Specker set and observation of its connection with the Greenberger-Horne-Zeilinger theorem,'' \pra \textbf{90}, 012119 (2014).

\bibitem{LeoCam}  L. J. Zhang, L. Neves, J. S. Lundeen, and I. A. Walmsley, ``A characterization of the single-photon sensitivity of an electron multiplying charge-coupled device,'' J. Phys. B \textbf{42}, 114011 (2009).

\bibitem{emccd2} E. Bolduc, D. Faccio, and J. Leach, ``Acquisition of multiple photon pairs with an EMCCD camera,'' J. Opt. \textbf{19}, 054006 (2017).

\bibitem{turb0} C. Gopaul and R. Andrews, ``The effect of atmospheric turbulence on entangled orbital angular momentum states,'' New J. Phys. \textbf{9}, 94 (2007).

\bibitem{turb1} A. K. Jha, G. A. Tyler, and R. W. Boyd, ``Effects of atmospheric turbulence on the entanglement of spatial two-qubit states,'' \pra \textbf{81}, 053832 (2010).

\bibitem{turb2} N. D. Leonhard, V. N. Shatokhin, and A. Buchleitner, ``Universal entanglement decay of photonic-orbital-angular-momentum qubit states in atmospheric turbulence,'' \pra \textbf{91}, 012345 (2015).

\bibitem{turb3} F. S. Roux, ``Biphoton states in correlated turbulence,'' \pra \textbf{95}, 023809 (2017).

\bibitem{Torres1} C. I. Osorio, G. Molina-Terriza, and J. P. Torres, ``Correlations in orbital angular momentum of spatially entangled paired photons generated in parametric down-conversion,'' \pra \textbf{77}, 015810 (2008).

\bibitem{OAMqd} J. Romero, D. Giovannini, S. Franke-Arnold, S. M. Barnett, and M. J. Padgett, ``Increasing the dimension in high-dimensional two-photon orbital angular momentum entanglement,'' \pra \textbf{86}, 012334 (2012).

\bibitem{OAMqd2} Y. Zhang, F. S. Roux, T. Konrad, M. Agnew, J. Leach, and A. Forbes, ``Engineering two-photon high-dimensional states through quantum interference,'' Sci. Adv. \textbf{2}, e150116 (2016).

\bibitem{OAMqd3} F. Wang, M. Erhard, A. Babazadeh, M. Malik, M. Krenn, and A. Zeilinger, ``Generation of the complete four-dimensional Bell basis,'' Optica \textbf{4}, 1462 (2017).

\bibitem{OAMZ} R. Fickler, G. Campbell, B. Buchler, P. K. Lam, and A. Zeilinger, ``Quantum entanglement of angular momentum states with quantum numbers up to 10,010,'' Proc. Natl. Acad. Sci. \textbf{113}, 13642 (2016).

\bibitem{SPDC}  C. H. Monken, P. H. S. Ribeiro, and S. Pádua, ``Transfer of angular spectrum and image formation in spontaneous parametric down-conversion,'' Phys. Rev. A 57, 3123 (1998).

\bibitem{tbo1}  E. S. Gómez, W. A. T. Nogueira, C. H. Monken, and G. Lima, ``Quantifying the non-Gaussianity of the state of spatially correlated down-converted photons,'' Opt. Express \textbf{20}, 3753 (2012).

\bibitem{tbo2}  E. S. Gómez, G. Cañas, E. Acuña, W. A. T. Nogueira, and G. Lima, ``Non-Gaussian-state generation certified using the Einstein-Podolsky-Rosen-steering inequality,'' \pra~\textbf{91}, 013801 (2015).

\bibitem{ExterNF2} H. D.-L. Pires and M. P. van Exter, ``Near-field correlations in the two-photon field,'' \pra~\textbf{80}, 053820 (2009).

\bibitem{cvtomo} A. I. Lvovsky and M. G. Raymer, ``Continuous-variable optical quantum-state tomography,'' \rmp \textbf{81}, 299 (2009).

\bibitem{reid} M. D. Reid, ``Demonstration of the Einstein-Podolsky-Rosen paradox using nondegenerate parametric amplification,'' Phys. Rev. A \textbf{40}, 913 (1989).

\bibitem{luan} L.-M. Duan, G. Giedke, J. I. Cirac, and P. Zoller, ``Inseparability criterion for continuous variable systems,'' \prl~\textbf{84}, 2722 (2000).

\bibitem{mancini} S. Mancini, V. Giovannetti, D. Vitali, and P. Tombesi, ``Entangling macroscopic oscillators exploiting radiation pressure,'' \prl \textbf{88}, 120401 (2002).

\bibitem{Eisert} M. Ostermeyer, D. Korn, D. Puhlmann, C. Henkel, and J. Eisert, ``Two-dimensional characterization of spatially entangled photon pairs,'' J. Mod. Opt. \textbf{56}, 1829-1837 (2009).

\bibitem{steveng} R. M. Gomes, A. Salles, F. Toscano, P. H. Souto Ribeiro, and S. P. Walborn, ``Quantum entanglement beyond Gaussian criteria,'' Proc. Natl. Acad. Sci. \textbf{106}, 21517-21520 (2009).

\bibitem{Gaussimp} J. Eisert, S. Scheel, and M. B. Plenio, ``Distilling Gaussian states with Gaussian operations is impossible,'' \prl~\textbf{89}, 137903 (2002).

\bibitem{Lizana2008} A. Lizana, I. Moreno, A. M\'arquez, C. Iemmi, E. Fern\'andez, J. Campos, and M. J. Yzuel, ``Time fluctuations of the phase modulation in a liquid crystal on silicon display: characterization and effects in diffractive optics,'' Opt. Express \textbf{16}, 16711 (2008).

\bibitem{QKDOAM1} M. Mafu, A. Dudley, S. Goyal, D. Giovannini, M. McLaren, M. J. Padgett, T. Konrad, F. Petruccione, N. Lütkenhaus, and A. Forbes, ``Higher-dimensional orbital-angular-momentum-based quantum key distribution with mutually unbiased bases,'' \pra \textbf{88}, 032305 (2013).

\bibitem{QKDOAM2} H. Lai, M.-X. Luo, J. Pieprzyk, J. Zhang, L. Pan, S. Li, and M. A. Orgun, ``Fast and simple high-capacity quantum cryptography with error detection,'' Sci. Rep. \textbf{7}, 46302 (2017).

\bibitem{Zhang06} J. Zhang and S. L. Braunstein, ``Continuous-variable Gaussian analog of cluster states,'' \pra~\textbf{73}, 032318 (2006).

\bibitem{Menicucci06} N. C. Menicucci, P. van Loock, M. Gu, C. Weedbrook, T. C. Ralph, and M. A. Nielsen, ``Universal quantum computation with continuous-variable cluster states,'' \prl~\textbf{97}, 110501 (2006).

\bibitem{Yukawa08} M. Yukawa, R. Ukai, P.van Loock, and A. Furusawa, ``Experimental generation of four-mode continuous-variable cluster states'', \pra~\textbf{78}, 012301 (2008).

\end{thebibliography}
\end{document}